\documentclass[preprint,prd,aps,showpacs,showkeys,nofootinbib]{revtex4}
\usepackage{graphicx}
\begin{document}

\title{Muon anomalous magnetic dipole moment in the $\mu\nu$SSM }

\author{Hai-Bin Zhang$^{a,b}$\footnote{Corresponding author.\\hbzhang@hbu.edu.cn},
Chang-Xin Liu$^{a,b}$\footnote{LIUchangxinZ@163.com},
Jin-Lei Yang$^{c,d}$\footnote{JLYangJL@163.com},
Tai-Fu Feng$^{a,b,e}$\footnote{fengtf@hbu.edu.cn}}

\affiliation{$^a$Department of Physics, Hebei University, Baoding, 071002, China\\
$^b$Key Laboratory of High-precision Computation and Application of Quantum Field Theory of Hebei Province, Baoding, 071002, China\\
$^c$CAS Key Laboratory of Theoretical Physics, Institute of Theoretical Physics, Chinese Academy of Sciences, Beijing 100190, China\\
$^d$School of Physical Sciences, University of Chinese Academy of Sciences, Beijing 100049, China\\
$^e$College of Physics, Chongqing University, Chongqing, 400044, China}

\begin{abstract}
Recently, the Muon g-2 experiment at Fermilab has measured the muon anomalous magnetic dipole moment (MDM), $a_\mu=(g_\mu-2)/2$, which reported that the new experimental average increases the tension between experiment and the standard model (SM) prediction to 4.2$\sigma$. In this work, we reanalyse the muon anomalous MDM at two-loop level in the $\mu$ from $\nu$ Supersymmetric Standard Model ($\mu\nu$SSM) combined with the updated experimental average. The $\mu\nu$SSM can explain the current tension between the experimental measurement and the SM theoretical prediction for the muon anomalous MDM, constrained by the 125 GeV Higgs boson mass and decays, the rare decay $\bar{B}\rightarrow X_s\gamma$ and so on. We also investigate the anomalous MDM of the electron and tau lepton,   $a_e=(g_e-2)/2$ and $a_\tau=(g_\tau-2)/2$, at two-loop level in the $\mu\nu$SSM. In addition, the 125 GeV Higgs boson decays to a pair of charged leptons in the $\mu\nu$SSM are also analysed.
\end{abstract}


\maketitle

\section{Introduction\label{sec1}}

The anomalous magnetic dipole moment (MDM) of the muon, $a_\mu=(g_\mu-2)/2$, recently has been measured by the Muon g-2 experiment at Fermilab \cite{MDM-exp1,MDM-exp2,MDM-exp3,MDM-exp4}, which reported that the result is 3.3 standard deviations ($\sigma$) greater than the standard model (SM) prediction based on its Run-1 data and is in agreement with the previous Brookhaven National Laboratory (BNL) E821 measurement \cite{muon-exp}. Combinated with previous E821 measurement, the new experimental average for the difference between the experimental measurement and the SM prediction \cite{Aoyama:2020ynm} of $a_\mu$ is given by
\begin{eqnarray}
\Delta a_{\mu}=a_{\mu}^{exp}-a_{\mu}^{SM}=(25.1\pm5.9)\times10^{-10},
\label{MDM-exp}
\end{eqnarray}
which increases the tension between  the experimental measurement and the SM theoretical prediction to 4.2$\sigma$. This result will further motivate the development of the SM extensions. There are tons of papers researched the muon anomalous MDM in the
Refs.~\cite{Jegerlehner:2009ry,Davier:2010nc,Eidelman:1995ny,Moroi:1995yh,Aoyama:2012wk,Hagiwara:2006jt,
Bijnens:1995xf,Martin:2001st,Lindner:2016bgg,Hagiwara:2003da,Bijnens:2001cq,Chattopadhyay:1995ae,Davier:2017zfy,
Bijnens:1995cc,Hagiwara:2002ma,Hayakawa:1995ps,Everett:2001tq,deRafael:1993za,Chattopadhyay:2001vx,
Hayakawa:1996ki,Hisano:2001qz,DeTroconiz:2001rip,Baek:2001kca,Davoudiasl:2012ig,RamseyMusolf:2002cy,Benayoun:2012wc,
Colangelo:2014qya,Komine:2001fz,Kinoshita:2005sm,Wang:2015kuj,Allanach:2015gkd,Belanger:2001am,
Chang:2000ii,Aoyama:2007dv,Passera:2006gc,Cao:2019evo,Abdughani:2019wai,Wang:2018vrr,Wang:2017vxj,
Kiritsis:2002aj,Padley:2015uma,Li:2018aov,Li:2020dbg,Cao:2021lmj,Chen:2021rnl,
Yin:2021yqy,Yin:2020afe,Sabatta:2019nfg,vonBuddenbrock:2019ajh,vonBuddenbrock:2016rmr,
Okada:2016wlm,Fukuyama:2016mqb,Belanger:2017vpq,Tran:2018kxv,
g-2muonQCD,g-2muon,g-2muon1,g-2muon2,g-2muon3,g-2muon4,g-2muon5,g-2muon6,g-2muon7,g-2muon8,g-2muon9,g-2muon10,
g-2muon11,g-2muon12,g-2muon13,g-2muon14,g-2muon15,g-2muon16,g-2muon17,g-2muon18,g-2muon19,g-2muon20,
g-2muon21,g-2muon22,g-2muon23,g-2muon24,g-2muon25,g-2muon26,g-2muon27,g-2muon28,g-2muon29,g-2muon30}
and references therein. However, it is worth mentioning that the latest result obtained by the lattice QCD calculation~\cite{Borsanyi:2020mff} of the leading order hadronic vacuum polarization contribution to $a_\mu$ is larger than the former result, which can accommodate the discrepancy between the experiment and the SM prediction, hence the discrepancy needs further scrutiny.

In this work, we will analyse the muon anomalous MDM at two-loop level in the $\mu$ from $\nu$ Supersymmetric Standard Model ($\mu\nu$SSM) \cite{mnSSM,mnSSM1,mnSSM1-1,mnSSM2,mnSSM2-1,Zhang1,Zhang2}, combined with the new experimental average. Through introducing three singlet right-handed neutrino superfields $\hat{\nu}_i^c$ ($i=1,2,3$), the $\mu$$\nu$SSM can solve the $\mu$ problem~\cite{m-problem} of the minimal supersymmetric standard model (MSSM)~\cite{MSSM,MSSM1,MSSM2,MSSM3,MSSM4}, and can generate three tiny neutrino masses through TeV scale seesaw mechanism
\cite{mnSSM1,neutrino-mass,neu-mass1,neu-mass2,neu-mass3,neu-mass4,neu-mass5,neu-mass6}.

The corresponding superpotential of the $\mu$$\nu$SSM is given as \cite{mnSSM,mnSSM1}
\begin{eqnarray}
&&W={\epsilon _{ab}}\left( {Y_{{u_{ij}}}}\hat H_u^b\hat Q_i^a\hat u_j^c + {Y_{{d_{ij}}}}\hat H_d^a\hat Q_i^b\hat d_j^c
+ {Y_{{e_{ij}}}}\hat H_d^a\hat L_i^b\hat e_j^c \right)  \nonumber\\
&&\hspace{0.95cm}
+ {\epsilon _{ab}}{Y_{{\nu _{ij}}}}\hat H_u^b\hat L_i^a\hat \nu _j^c -  {\epsilon _{ab}}{\lambda _i}\hat \nu _i^c\hat H_d^a\hat H_u^b + \frac{1}{3}{\kappa _{ijk}}\hat \nu _i^c\hat \nu _j^c\hat \nu _k^c.
\label{eq-W}
\end{eqnarray}
where $a,b=1,2$ are SU(2) indices with antisymmetric tensor $\epsilon_{12}=1$, and $i,j,k=1,2,3$ are generation indices. The repeating indices imply the summation convention in the following. $Y_{u,d,e,\nu}$, $\lambda$, and $\kappa$ are dimensionless matrices, a vector, and a symmetric tensor.
In the superpotential, the effective bilinear terms $\epsilon _{ab} \varepsilon_i \hat H_u^b\hat L_i^a$ and $\epsilon _{ab} \mu \hat H_d^a\hat H_u^b$ can be generated, with $\varepsilon_i= Y_{\nu _{ij}} \left\langle {\tilde \nu _j^c} \right\rangle$ and $\mu  = {\lambda _i}\left\langle {\tilde \nu _i^c} \right\rangle$,  once the electroweak symmetry is broken. In the $\mu\nu$SSM, the general soft SUSY-breaking terms, the usual $D$- and $F$-term contributions of the tree-level scalar potential, and the mass matrices of the particles can be seen in Refs.~\cite{mnSSM1,mnSSM1-1,Zhang1}.
In the $\mu$$\nu$SSM, the gravitino or the axino can be the dark matter candidates~\cite{mnSSM1,mnSSM1-1,neu-mass3,DM1,DM2,DM3,DM4,DM5,DM6}.

In our previous work, the Higgs boson  mass and decay modes $h\rightarrow\gamma\gamma$, $h\rightarrow VV^*$ ($V=Z,W$), $h\rightarrow f\bar{f}$ ($f=b,\tau$), and $h\rightarrow Z\gamma$ in the $\mu\nu$SSM have been researched~\cite{Zhang-MASS,hrr,hLFV,hZr,hMZ}. Constrained by the 125 GeV Higgs boson mass and decays, here we will investigate the anomalous MDM of the charged leptons at two-loop level in the $\mu\nu$SSM, combined with the updated experimental average of the muon MDM.
For the electron anomalous MDM, $a_e=(g_e-2)/2$, the experimental result showed a negative $\sim2.4\sigma$ discrepancy between the measured value~\cite{Hanneke:2008tm} and the SM prediction~\cite{Parker:2018vye}. However, a new determination of the fine structure constant with a higher accuracy~\cite{Morel:2020}, obtained from the measurement of the recoil velocity on rubidium atoms, resulted into a re-evaluation of $a_e$ in the SM, bringing to a positive $\sim1.6\sigma$ discrepancy
\begin{eqnarray}
&&\bigtriangleup a_e\equiv a_e^{exp}-a_e^{SM}=(4.8\pm3.0)\times10^{-13}.
\label{MDMe-exp}
\end{eqnarray}
Interestingly, Now $\bigtriangleup a_e$ and $\bigtriangleup a_\mu$ are all positive.

Now, the measured averages of the signal strengths for the 125 GeV Higgs boson decays into two taus and bottom quarks relative to the standard model (SM) prediction are respectively $1.15^{+0.16}_{-0.15}$ and $1.04\pm0.13$ with high experimental precision \cite{PDG1}.
Although the Higgs boson decays to a pair of fermions of the third generation are now measured accurately by the Large Hadron Collider (LHC), the Higgs boson decays to a pair of fermions of the first or second generation are challenging to measure, due that the Yukawa couplings of the 125 GeV Higgs boson to fermions of the first and second generation are small than that of the third generation. However, the ATLAS and CMS Collaborations recently measured the 125 GeV Higgs boson decay to a pair of muons $h\rightarrow \mu \bar{\mu}$, which reported the signal strength relative to the SM prediction is $1.2\pm0.6$ with 2.0$\sigma$ \cite{ATLAS-h2u} and $1.19^{+0.40+0.15}_{-0.39-0.14}$ with 3.0$\sigma$ \cite{CMS-h2u}, respectively. The dimuon decay of the 125 GeV Higgs boson  $h\rightarrow \mu \bar{\mu}$ offers the best opportunity to measure the Higgs interactions with the second-generation fermions at the LHC. Within various theoretical frameworks, the 125 GeV Higgs boson decay $h\rightarrow \mu \bar{\mu}$ has been discussed \cite{Huu-M1,Huu-M2,Huu-M3,Huu-M4,Huu-M5,Huu-M6,Huu-M7,Huu-M8,Huu-M9,Huu-M10,Huu-M11,Huu-M12,Huu-M13}. Here, we will investigate the 125 GeV Higgs boson decay $h\rightarrow \mu \bar{\mu}$ at one-loop level in the $\mu\nu$SSM.

In the following, we briefly introduce the MDM of the charged leptons in Sec.~\ref{sec2}.  In Sec.~\ref{sec-h}, we give the decay width of the 125 GeV Higgs boson decays to a pair of charged leptons $h\rightarrow l_i \bar{l}_i$ at one-loop level. Sec.~\ref{sec-Num} and Sec.~\ref{sec-Sum} respectively show the numerical analysis and summary.

\section{The MDM of the charged leptons \label{sec2}}

\begin{figure}
\setlength{\unitlength}{1mm}
\centering
\begin{minipage}[c]{1.0\textwidth}
\includegraphics[width=4.2in]{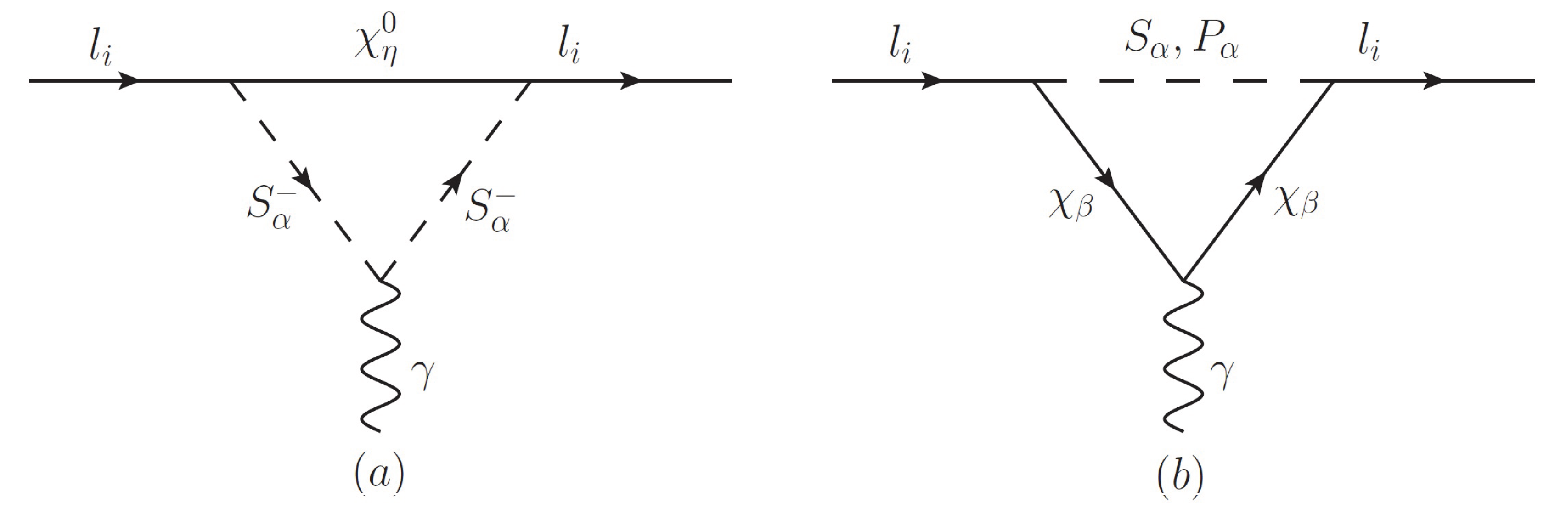}
\end{minipage}
\caption[]{The dominant one-loop diagrams, represent the contributions from neutral fermions $\chi_\eta^0$ and charged scalars $S_\alpha^-$ loops (a), and the contributions from charged fermions $\chi_\beta$ and neutral scalars $S_\alpha$ (or $P_\alpha$) loops (b).}
\label{Muon1}
\end{figure}

\begin{figure}
\setlength{\unitlength}{1mm}
\centering
\begin{minipage}[c]{1.0\textwidth}
\includegraphics[width=6.0in]{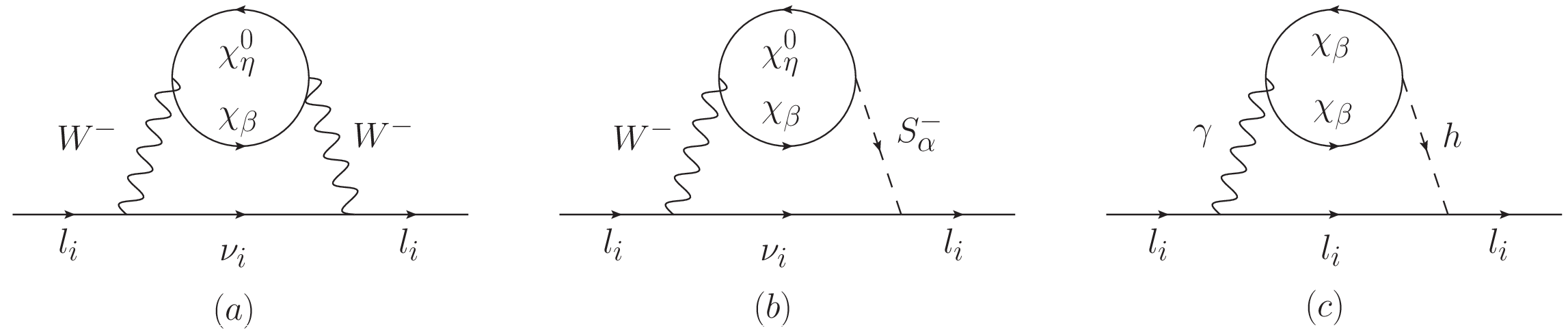}
\end{minipage}
\caption[]{The main two-loop rainbow diagram (a) and Barr-Zee type diagrams (b,c).}
\label{Muon2}
\end{figure}

The MDM of the charged leptons in the $\mu\nu$SSM can be written by the effective Lagrangian
\begin{eqnarray}
&&\mathcal{L}_{MDM}={e\over 4m_{l_i}}a_{l_i}\overline{l}_{i}\sigma^{\alpha \beta}l_{i}F_{\alpha \beta},
\end{eqnarray}
where $l_i$ represents the charged leptons which is on-shell, $m_{l_i}$ is the mass of the charged leptons, $\sigma^{\alpha\beta}=\frac{i}{2}[\gamma^\alpha,\gamma^\beta]$, $F_{\alpha\beta}$ denotes the electromagnetic field strength and the MDM of the charged leptons is $a_{l_i}=\frac{1}{2}(g_{l_i}-2)$.
Including main two-loop electroweak corrections, the MDM of the charged leptons in the $\mu\nu$SSM can be given by
\begin{eqnarray}
&&a_{l_i}^{\rm{SUSY}}=a_{l_i}^{\rm{one-loop}}+a_{l_i}^{\rm{two-loop}},
\end{eqnarray}
where the one-loop corrections $a_{l_i}^{\rm{one-loop}}$ are pictured in Fig.~\ref{Muon1} and  the main two-loop corrections $a_{l_i}^{\rm{two-loop}}$ are showed in Fig.~\ref{Muon2}.

In Fig.~\ref{Muon1}, the contributions to the MDM of the charged leptons at the one-loop level in  the $\mu\nu$SSM come from the neutral fermions  and charged scalars loops (neutral fermions $\chi_{\eta}^{0}$ and charged scalars $S_\alpha^-$  are loop particles) and the charged fermions and neutral scalars loop (charged fermions $\chi_{\beta}$ and neutral scalars $N_\alpha=S_\alpha,P_\alpha$  are loop particles). The concrete expressions of the one-loop corrections $a_{l_i}^{\rm{one-loop}}$ can be found in our previous related work~\cite{hLFV}, where one can use the charged leptons ${l_i}$  replace the muon ${l_\mu}$.

Here, the dominant contribution of the muon MDM $a_\mu$ comes from the charged fermions and neutral scalars loops in Fig.~\ref{Muon1}(b).
We check that the one-loop correction in the $\mu\nu$SSM is approximately in agreement with the MSSM and the Next-to-Minimal Supersymmetric Standard Model (NMSSM)~\cite{Moroi:1995yh,g-2muon2,g-2muon20}. Although the MDM of the muon in the $\mu\nu$SSM has roughly the same properties in the MSSM and NMSSM, it is subject to significantly relaxed limitations for the parameter space in the $\mu\nu$SSM  if researching the other physical quantities. Of course, through introducing three singlet right-handed neutrino superfields $\hat{\nu}_i^c$ ($i=1,2,3$) to solve the $\mu$ problem of the MSSM and generate three tiny neutrino masses, the $\mu$$\nu$SSM still can give some additional contributions to the muon MDM $a_\mu$ beyond the MSSM.

In Fig.~\ref{Muon2}, the main two-loop rainbow diagram (a) and Barr-Zee type diagrams (b,c) of $a_{l_i}$ in the $\mu\nu$SSM are showed,  which a closed fermion loop is attached to the virtual gauge bosons or scalars, and the corresponding corrections  to  $a_{l_i}$ are obtained by attaching a photon on all possible ways to the internal particles. In our previous work~\cite{hZr}, we show the main two-loop contributions of the muon MDM in the approximation $m_{\chi_{\eta}^{0}}\simeq m_{\chi_{\beta}}$. In this paper, we give the main two-loop contributions $a_{l_i}^{\rm{two-loop}}$ in the general case.

In the $\mu\nu$SSM, the main SUSY two-loop corrections of the MDM of the charged leptons can be given as
\begin{eqnarray}
a_{l_i}^{\rm{two-loop}}=a_{l_i}^{WW}+a_{l_{i}}^{WS}+a_{l_i}^{\gamma h},
\end{eqnarray}
where the terms $a_{l_i}^{WW},a_{l_i}^{WS},a_{l_i}^{\gamma h}$ are the contributions corresponding to Fig.~\ref{Muon2} (a-c). The contribution from the main two-loop rainbow diagram in Fig.~\ref{Muon2} (a) can be written by
\begin{eqnarray}
&&a_{l_i}^{\rm WW}={G_{_F}m_{l_i}^2\over 8\sqrt{2}\pi^4}
\Big\{\Big(|C_{L}^{W \overline{\chi}_{\eta}^{0}\chi_{\beta}}|^2+|C_{R}^{W\overline{\chi}_{\eta}^{0}\chi_{\beta}}|^2\Big)
T_1(1,x_{{\chi_{\eta}^{0}}},x{_{\chi_\beta}})\nonumber\\
&&\hspace{1.3cm}
+\Big(|C_{L}^{W\overline{\chi}_{\eta}^{0}\chi_{\beta}}|^2-|C_{R}^{W\overline{\chi}_{\eta}^{0}\chi_{\beta}}|^2\Big)
T_2(1,x_{{\chi_{\eta}^{0}}},x_{{\chi_\beta}})
\nonumber\\
&&\hspace{1.3cm}
+2(x_{{\chi_{\eta}^{0}}}x_{{\chi_\beta}})^{1/2}\Re(C_{R}^{W\overline{\chi}_{\eta}^{0}\chi_{\beta}*}
C_{L}^{W\overline{\chi}_{\eta}^{0}\chi_{\beta}})T_3(1,x_{{\chi_{\eta}^{0}}},x{_{\chi_\beta}})\Big\}\;,
\end{eqnarray}
with $x_i=m_i^2/m_W^2$. The expressions of form factors $T_i$ can be found in Refs. \cite{Feng3,Feng1,Feng2}.
Here,  the concrete expressions for couplings $C$ in the $\mu\nu$SSM can be seen in Ref.~\cite{Zhang1}.

The contribution from the main two-loop Barr-Zee type diagram in Fig.~\ref{Muon2} (b) can be given by
\begin{eqnarray}
&&a_{l_{i}}^{WS}={G_{_F}m_{l_{i}}m_{W}\over128\pi^4g_2}\Re{( C_{L}^{S_{\alpha}^{-}\overline{l}_i{\chi_{7+i}^0}} )}\nonumber\\
&&\hspace{1.3cm}
\times\Bigg\{(x_{{\chi_\beta}})^{1/2}
F_1(1,x_{S_{\alpha}^{-}},x_{{\chi_{\eta}^{0}}},x_{{\chi_\beta}})
\Re\Big(C_{L}^{S_{\alpha}^{-}\overline\chi_{\beta}{\chi}_{\eta}^{0}}C_{L}^{W \overline{\chi}_{\eta}^{0}\chi_{\beta}}+C_{R}^{S_{\alpha}^{-}\overline\chi_{\beta}{\chi}_{\eta}^{0}}C_{R}^{W \overline{\chi}_{\eta}^{0}\chi_{\beta}}\Big)
\nonumber\\
&&\hspace{1.3cm}
+(x_{{\chi_{\eta}^{0}}})^{1/2}
F_2(1,x_{{S_{\alpha}^{-}}},x_{{\chi_{\eta}^{0}}},x_{{\chi_\beta}})
\Re\Big(C_{L}^{S_{\alpha}^{-}\overline\chi_{\beta}{\chi}_{\eta}^{0}}C_{R}^{W \overline{\chi}_{\eta}^{0}\chi_{\beta}}+C_{R}^{S_{\alpha}^{-}\overline\chi_{\beta}{\chi}_{\eta}^{0}}C_{L}^{W \overline{\chi}_{\eta}^{0}\chi_{\beta}}\Big)
\nonumber\\
&&\hspace{1.3cm}
+(x_{{\chi_\beta}})^{1/2}
F_3(1,x_{S_{\alpha}^{-}},x_{{\chi_{\eta}^{0}}},x_{{\chi_\beta}})
\Re\Big(C_{L}^{S_{\alpha}^{-}\overline\chi_{\beta}{\chi}_{\eta}^{0}}C_{L}^{W \overline{\chi}_{\eta}^{0}\chi_{\beta}}-C_{R}^{S_{\alpha}^{-}\overline\chi_{\beta}{\chi}_{\eta}^{0}}C_{R}^{W \overline{\chi}_{\eta}^{0}\chi_{\beta}}\Big)
\nonumber\\
&&\hspace{1.3cm}
+(x_{{\chi_{\eta}^{0}}})^{1/2}
F_4(1,x_{S_{\alpha}^{-}},x_{{\chi_{\eta}^{0}}},x_{{\chi_\beta}})
\Re\Big(C_{L}^{S_{\alpha}^{-}\overline\chi_{\beta}{\chi}_{\eta}^{0}}C_{R}^{W \overline{\chi}_{\eta}^{0}\chi_{\beta}}-C_{R}^{S_{\alpha}^{-}\overline\chi_{\beta}{\chi}_{\eta}^{0}}C_{L}^{W \overline{\chi}_{\eta}^{0}\chi_{\beta}}\Big)\Bigg\}\;,
\end{eqnarray}
where the expressions of form factors $F_{i}$ can be seen in Ref. \cite{Feng3}.
Here, in the $\mu\nu$SSM,  $h$ denotes $S_1$, $l_i$ is denoted to ${\chi}_{2+i}$. Considered that the masses of the charged scalars $S_{\alpha}^{-}$ are more larger than the mass of $W$ gauge boson constrained by the present experiments, the contribution from the main two-loop Barr-Zee type diagram $a_{l_{i}}^{WS}$ is more smaller than the contribution from the main two-loop rainbow diagram $a_{l_i}^{\rm WW}$.

The contribution from the main two-loop Barr-Zee type diagram in Fig.~\ref{Muon2} (c) can be  written  as
\begin{eqnarray}
a_{l_{i}}^{\gamma h}={{-G_{_F} m_{l_i}m_{W} s_W^2}
\over {16\pi^4}}(x_{{\chi_\beta}})^{1/2} T_{11}(x_{{h}},x_{{\chi_{\beta}}},x_{{\chi_{\beta}}})\Re(C_{L}^{h \overline{l}_i {l}_{i}} C_{L}^{h\overline{\chi}_{\beta}{\chi}_{\beta}})\;.
\end{eqnarray}
Through the numerical calculation, normalized to the one-loop corrections $a_{l_i}^{\rm{one-loop}}$, the two-loop corrections of the MDM $a_{l_i}^{\rm{two-loop}}$ in the $\mu\nu$SSM may reach about 10$\%$, when $\tan\beta$ is large and  the masses of superpartners are small constrained from the experiments. So, one-loop correction alone is sufficient to explain the g-2 of the muon and satisfy other experimental constraints. In the following numerical analysis, the two-loop corrections of the muon MDM are still considered to be more precise.

\section{$h\rightarrow l_i \bar{l}_i$ in the $\mu\nu$SSM\label{sec-h}}

The corresponding effective amplitude for 125 GeV Higgs decay $h\rightarrow l_i \bar{l}_i$ can be written as
\begin{eqnarray}
\mathcal{M}= h{\bar l_i}({F_L^{i}}{P_L} + {F_R^{i}}{P_R}){l_i}.
\end{eqnarray}
The decay width of $h\rightarrow l_i\bar{l}_i$ can be obtained as
\begin{eqnarray}
{\Gamma}(h\rightarrow l_i\bar{l}_i) \simeq \frac{m_h}{16\pi}\Big({\left| {F_L^{i}} \right|^2} + {\left| {F_R^{i}} \right|^2}\Big).
\end{eqnarray}

The contribution from the tree level in the $\mu\nu$SSM can be written as
\begin{eqnarray}
F_{L}^{(tree)i}=F_{R}^{(tree)i}  = \frac{m_{l_i}}{\sqrt{2}\upsilon\cos \beta}R_{S_{11}},
\label{FLtree}
\end{eqnarray}
where $m_{l_i}$  denotes the mass of the lepton $l_i$, $\upsilon\simeq174 $GeV, $R_S$ is the unitary matrix which diagonalizes the mass matrix of CP-even neutral scalars~\cite{Zhang-MASS}. For the SM, the contribution from the tree level can be written by
\begin{eqnarray}
F_{L(\rm{SM})}^{(tree)i}=F_{R(\rm{SM})}^{(tree)i}  = \frac{m_{l_i}}{\sqrt{2}\upsilon}.
\end{eqnarray}
The running lepton masses $m_{l_i}(\Lambda)$ are related to the pole masses $m_{l_i}$ through~\cite{RG}
\begin{eqnarray}
m_{l_i}(\Lambda)=m_{l_i}\Big\{ 1-\frac{\alpha(\Lambda)}{\pi} \Big[1+\frac{3}{4} \ln \frac{\Lambda^2}{m_{l_i}^2}\Big]\Big\}.
\end{eqnarray}

Similarly to the decays $h\rightarrow l_i\bar{l}_i$, the decay width of the 125 GeV Higgs decay to the down-type quarks $h\rightarrow d_i \bar{d}_i$ can be given as
\begin{eqnarray}
{\Gamma}(h\rightarrow d_i \bar{d}_i) \simeq \frac{N_c m_h}{16\pi}\Big({\left| {F_{dL}^{i}} \right|^2} + {\left| {F_{dR}^{i}} \right|^2}\Big),
\end{eqnarray}
with $N_c=3$, and the tree level contribution in the $\mu\nu$SSM is
\begin{eqnarray}
F_{dL}^{(tree)i}=F_{dR}^{(tree)i}  = \frac{m_{d_i}}{\sqrt{2}\upsilon\cos \beta}R_{S_{11}},
\end{eqnarray}
where $m_{d_i}$  denotes the mass of the down-type quarks $d_i$. For the SM, the contribution from the tree level can be written by
\begin{eqnarray}
F_{dL(\rm{SM})}^{(tree)i}=F_{dR(\rm{SM})}^{(tree)i}  = \frac{m_{d_i}}{\sqrt{2}\upsilon}.
\end{eqnarray}

The difference of the decay width of $h\rightarrow f_i \bar{f}_i$ of the $\mu\nu$SSM (${\Gamma}_{\rm{NP}}(h\rightarrow f_i \bar{f}_i)$) and that of the SM (${\Gamma}_{\rm{SM}}(h\rightarrow f_i \bar{f}_i)$) in the tree level can be given as
\begin{eqnarray}
\delta_{tree}\equiv {{\Gamma}_{\rm{NP}}(h\rightarrow f_i \bar{f}_i)-{\Gamma}_{\rm{SM}}(h\rightarrow f_i \bar{f}_i)\over {\Gamma}_{\rm{SM}}(h\rightarrow f_i \bar{f}_i)}=\frac{R_{S_{11}}^2}{\cos^2 \beta}-1.
\label{eq-tree}
\end{eqnarray}
Here $f_i=l_i,\,d_i$, due that the tree-level contribution of the Higgs boson decay into leptons is identical for the Higgs boson decay into down-tpye quarks. The numerical results  can show that  the  ratio $\delta_{tree}$ can be about $1\%$, when the parameter $\tan\beta$ in the $\mu\nu$SSM is small.

\begin{figure}
\begin{center}
\begin{minipage}[c]{0.5\textwidth}
\includegraphics[width=2.2in]{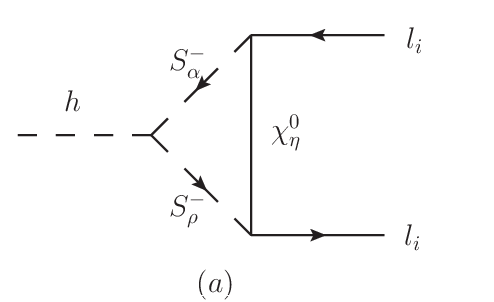}
\end{minipage}
\begin{minipage}[c]{0.39\textwidth}
\includegraphics[width=2.2in]{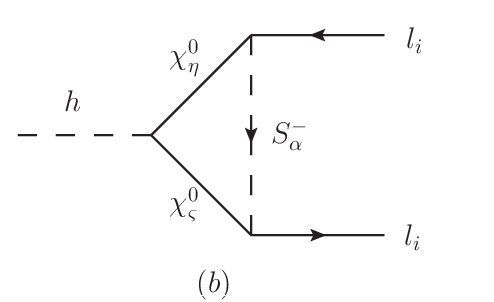}
\end{minipage}\\
\begin{minipage}[c]{0.5\textwidth}
\includegraphics[width=2.2in]{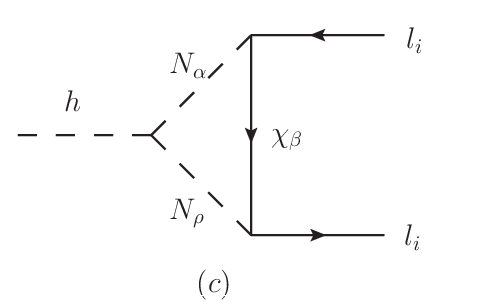}
\end{minipage}
\begin{minipage}[c]{0.39\textwidth}
\includegraphics[width=2.2in]{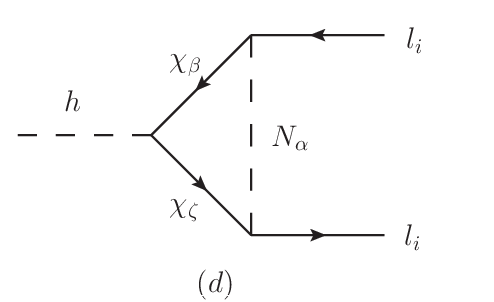}
\end{minipage}
\caption[]{The one-loop diagrams for $h\rightarrow l_i\bar{l}_i$ in the $\mu\nu$SSM. (a,b) represent the contributions from charged scalar $S_{\alpha,\rho}^-$ and neutral fermion $\chi_{\eta,\varsigma}^0$ loops, while (c,d) represent the contributions from neutral scalar $N_{\alpha,\rho}$ ($N=S,P$) and charged fermion $\chi_{\beta,\zeta}$ loops. }
\label{figh}
\end{center}
\end{figure}

The one-loop electroweak correction for $h\rightarrow l_i\bar{l}_i$ in the SM is approximated by\cite{Huu-M3,Huu-M4,Huu-M5,Huu-M6,Huu-M7}
\begin{eqnarray}
{\Gamma}_{SM}^{(one)}(h\rightarrow l_i\bar{l}_i) \simeq {\Gamma}_{SM}^{(tree)}(h\rightarrow l_i\bar{l}_i)\delta_{week}^{l},
\end{eqnarray}
with
\begin{eqnarray}
\delta_{week}^{l}=\frac{G_F}{8\pi^2 \sqrt{2}}\Big[  7m_t^2 + m_W^2\Big(-5 +  \frac{3\log{c_W^2}}{s_W^2}
 \Big) - m_Z^2 \frac{6(1-8s_W^2+16s_W^4)-1}{2} \Big],
\end{eqnarray}
where the contributions come from $t$ quark, $W$ boson and $Z$ boson. The numerical result shows that the one-loop electroweak contribution relative to the tree contribution $\delta_{week}^{l}$ is about 1.7\%.

The one-loop diagrams for $h\rightarrow l_i\bar{l}_i$ in the $\mu\nu$SSM beyond  the SM are depicted by Fig.~\ref{figh}. Then, the contribution from the one-loop diagrams in the $\mu\nu$SSM can be written by
\begin{eqnarray}
F_{L,R}^{(one)i} = F_{L,R}^{(a)i} + F_{L,R}^{(b)i} + F_{L,R}^{(c)i} + F_{L,R}^{(d)i},
\end{eqnarray}
where $F_{L,R}^{(a,b)i}$ denotes the contributions from charged scalar $S_{\alpha,\rho}^-$ and neutral fermion $\chi_{\eta,\varsigma}^0$ (upper index 0 shows neutral) loops, and $F_{L,R}^{(c,d)i}$ stands for the contributions from the neutral scalar $N_{\alpha,\rho}$ ($N=S,P$) and charged fermion $\chi_{\beta,\zeta}$ loops, respectively.

After integrating the heavy freedoms out, we formulate the neutral fermion loop contributions $F_{L,R}^{(a,b)i}$ as follows:
\begin{eqnarray}
&&F_L^{(a)i} =  \frac{{m_{{\chi _\eta ^0}}}{C^{S^\pm}_{1 \alpha \rho }}}{{m_W^2}}
C_L^{S_\rho ^ - {{\bar l }_{i}} \chi _\eta ^0}
C_L^{S_\alpha ^{-\ast} \bar \chi _\eta ^0 {l_{i}}}
{G_1}({x_{\chi _\eta ^0}},{x_{S_\alpha ^ - }},{x_{S_\rho ^ - }}) ,\nonumber\\
&&F_L^{(b)i} =  \frac{{m_{{\chi _\varsigma^0 }}}{m_{{\chi _\eta^0 }}}}{{m_W^2}}
C_L^{{S_\alpha^- }{{\bar l}_{i}}{\chi _\varsigma^0 }}C_L^{h{{\bar \chi }_\varsigma^0 }{\chi _\eta^0 }}
C_L^{{S_\alpha^{-\ast} }{{\bar \chi }_\eta^0}{l_{i}}}{G_1}({x_{{S_\alpha^- }}},{x_{{\chi _\varsigma^0 }}},{x_{{\chi _\eta^0 }}})\nonumber\\
&&\hspace{1.3cm} + \:  C_L^{{S_\alpha^- }{{\bar l}_{i}}{\chi _\varsigma^0 }}C_R^{h{{\bar \chi }_\varsigma^0 }{\chi _\eta^0 }}
C_L^{{S_\alpha^{-\ast} }{{\bar \chi }_\eta^0}{l_{i}}}{G_2}({x_{{S_\alpha^- }}},{x_{{\chi _\varsigma^0 }}},{x_{{\chi _\eta^0 }}}) ,\nonumber\\
&&F_R^{(a,b)i} = \left. {F_L^{(a,b)i}} \right|{ _{L \leftrightarrow R}} .
\end{eqnarray}
Here, the concrete expressions for the couplings $C$ can be found in Refs.~\cite{hrr,hLFV}, and the loop functions $G_{i}$ are given as
\begin{eqnarray}
&&{G_1}({\textit{x}_1 , \textit{x}_2 , \textit{x}_3}) =  \frac{1}{{16{\pi ^2}}}\Big[ \frac{{{x_1}\ln {x_1}}}{{({x_1} - {x_2})({x_1} - {x_3})}}
+ \frac{{{x_2}\ln {x_2}}}{{({x_2} - {x_1})({x_2} - {x_3})}} \nonumber\\
&&\hspace{2.9cm} + \frac{{{x_3}\ln {x_3}}}{{({x_3} - {x_1})({x_3} - {x_2})}}\Big], \\
&&{G_2}({\textit{x}_1 , \textit{x}_2 , \textit{x}_3}) =  \frac{1}{{16{\pi ^2}}}\Big[  \frac{{x_1^2\ln {x_1}}}{{({x_1} - {x_2})({x_1} - {x_3})}}
+ \frac{{x_2^2\ln {x_2}}}{{({x_2} - {x_1})({x_2} - {x_3})}} \nonumber\\
&&\hspace{2.9cm}  + \frac{{x_3^2\ln {x_3}}}{{({x_3} - {x_1})({x_3} - {x_2})}} \Big].\quad\;\;
\end{eqnarray}

In a similar way, the charged fermion loop contributions $F_{L,R}^{(c,d)i}$ are
\begin{eqnarray}
&&F_L^{(c)i} = \sum\limits_{N=S,P} \frac{{m_{{\chi _\beta }}}{C^{N}_{1 \alpha \rho }}}{{m_W^2}}
C_L^{N_\rho  {{\bar l }_{i}}\chi _\beta }
C_L^{N_\alpha \bar \chi _\beta {l_{i}} }
{G_1}({x_{\chi _\beta }},{x_{N_\alpha  }},{x_{N_\rho  }}) ,\nonumber\\
&&F_L^{(d)i} = \sum\limits_{N=S,P} \Big[ C_L^{{N_\alpha }{{\bar l}_{i}}{\chi _\zeta }}C_R^{h{{\bar \chi }_\zeta }{\chi _\beta }}C_L^{{N_\alpha }{{\bar \chi }_\beta }{l_{i}}}
{G_2}({x_{{N_\alpha }}},{x_{{\chi _\zeta }}},{x_{{\chi _\beta }}})\nonumber\\
&&\hspace{1.4cm} + \frac{{m_{{\chi _\zeta }}}{m_{{\chi _\beta }}}}{{m_W^2}}
C_L^{{N_\alpha }{{\bar l}_{i}}{\chi _\zeta }}C_L^{h{{\bar \chi }_\zeta }{\chi _\beta }}
C_L^{{N_\alpha }{{\bar \chi }_\beta }{l_{i}}}{G_1}({x_{{N_\alpha }}},{x_{{\chi _\zeta }}},{x_{{\chi _\beta }}})  \Big],\nonumber\\
&&F_R^{(c,d)i} = \left. {F_L^{(c,d)i}} \right|{ _{L \leftrightarrow R}} .
\end{eqnarray}

\section{Numerical analysis\label{sec-Num}}

Firstly, we take some appropriate parameter space in the $\mu\nu$SSM. For soft SUSY-breaking mass squared parameters, we make the minimal flavor violation (MFV) assumptions
\begin{eqnarray}
&&\hspace{-0.9cm}m_{\tilde Q_{ij}}^2 = m_{{{\tilde Q_i}}}^2{\delta _{ij}}, \quad
m_{\tilde u_{ij}^c}^2 = m_{{{\tilde u_i}^c}}^2{\delta _{ij}}, \quad
m_{\tilde d_{ij}^c}^2 = m_{{{\tilde d_i}^c}}^2{\delta _{ij}}, \nonumber\\
&&\hspace{-0.9cm}m_{{{\tilde L}_{ij}}}^2 = m_{{\tilde L}}^2{\delta _{ij}}, \quad
m_{\tilde e_{ij}^c}^2 = m_{{{\tilde e}^c}}^2{\delta _{ij}}, \quad
m_{\tilde \nu_{ij}^c}^2 = m_{\tilde \nu_{i}^c}^2{\delta _{ij}},
\label{MFV}
\end{eqnarray}
where $i,\;j,\;k =1,\;2,\;3 $. $m_{\tilde \nu_i^c}^2$ can be constrained by the minimization conditions of the neutral scalar potential seen in Ref.~\cite{Zhang-MASS}. For some coupling  parameters, we also choose the MFV assumptions
\begin{eqnarray}
&&\hspace{-0.9cm}{\kappa _{ijk}} = \kappa {\delta _{ij}}{\delta _{jk}}, \quad
{({A_\kappa }\kappa )_{ijk}} =
{A_\kappa }\kappa {\delta _{ij}}{\delta _{jk}}, \quad
\upsilon_{\nu_i^c}=\upsilon_{\nu^c}, \quad
\lambda _i = \lambda,\quad {({A_\lambda }\lambda )}_i = {A_\lambda }\lambda, \nonumber\\
&&\hspace{-0.9cm}
{Y_{{e_{ij}}}} = {Y_{{e_i}}}{\delta _{ij}},\quad
{({A_e}{Y_e})_{ij}} = {A_{e}}{Y_{{e_i}}}{\delta _{ij}},\quad
{Y_{{\nu _{ij}}}} = {Y_{{\nu _i}}}{\delta _{ij}},\quad
(A_\nu Y_\nu)_{ij}={a_{{\nu_i}}}{\delta _{ij}},
\label{MFV1}
\end{eqnarray}
Through our previous work~\cite{neu-mass6}, we have discussed in detail how the neutrino oscillation data constrain left-handed sneutrino VEVs $\upsilon_{\nu_i} \sim \mathcal{O}(10^{-4}\,{\rm{GeV}})$ and neutrino Yukawa couplings $Y_{\nu_i} \sim \mathcal{O}(10^{-7})$ in the $\mu\nu$SSM via the TeV scale seesaw mechanism. In the following, we choose $m_{\nu_1} =10^{-2}$ eV as the lightest neutrino and assume the neutrino mass spectrum with normal ordering, using neutrino oscillation experimental data \cite{PDG1} to constrain the parameters $\upsilon_{\nu_i}$ and $Y_{\nu_i}$.
Considered experimental data on quark mixing, one can have
\begin{eqnarray}
&&\hspace{-0.75cm}\;\,{Y_{{u _{ij}}}} = {Y_{{u _i}}}{V_{L_{ij}}^u},\quad
 (A_u Y_u)_{ij}={A_{u_i}}{Y_{{u_{ij}}}},\nonumber\\
&&\hspace{-0.75cm}\;\,{Y_{{d_{ij}}}} = {Y_{{d_i}}}{V_{L_{ij}}^d},\quad
(A_d Y_d)_{ij}={A_{d}}{Y_{{d_{ij}}}},
\end{eqnarray}
and $V=V_L^u V_L^{d\dag}$ denotes the CKM matrix.
\begin{eqnarray}
{Y_{{u_i}}} = \frac{{{m_{{u_i}}}}}{{{\upsilon_u}}},\qquad {Y_{{d_i}}} = \frac{{{m_{{d_i}}}}}{{{\upsilon_d}}},\qquad {Y_{{e_i}}} = \frac{{{m_{{l_i}}}}}{{{\upsilon_d}}},
\end{eqnarray}
where the $m_{u_{i}},m_{d_{i}}$ and $m_{l_{i}}$ stand for up-quark, down-quark and charged lepton masses.

Through analysis of the parameter space of the $\mu\nu$SSM \cite{mnSSM1}, we can choose reasonable parameter values to  be $\kappa=0.4$, ${A_{\kappa}}=-300\;{\rm GeV}$, $\lambda=0.1$, $A_\lambda=500\;{\rm GeV}$ and $A_{u_{1,2}}=A_{d}=A_{e}=1\;{\rm TeV}$ for simplicity. Considering the direct search for supersymmetric particles~\cite{PDG1},  we take  $m_{{\tilde Q}_{1,2,3}}=m_{{\tilde u_{1,2}}^{c}}=m_{{\tilde d_{1,2,3}}^{c}}=2\;{\rm TeV}$, $M_3=2.5\;{\rm TeV}$. For simplicity, we will choose the gauginos' Majorana masses $M_{1}=M_2$. As key parameters, $A_{u_{3}}\equiv A_t$, $m_{{\tilde u}^c_3}$ and $\tan\beta \equiv \upsilon_u/\upsilon_d$ greatly affect the lightest Higgs boson mass. Therefore, the free parameters that affect our next analysis are
\begin{eqnarray}
\tan \beta ,\quad \upsilon_{\nu^c}, \quad M_2, \quad m_{{\tilde L}}, \quad m_{{{\tilde e}^c}},\quad m_{{\tilde u}^c_3}, \quad A_t.
\end{eqnarray}

\begin{table}
\begin{tabular}{|cccc|}
\hline
Parameters&Min&Max&\\
\hline
$\tan \beta$&4&40&\\
$v_{\nu^{c}}/{\rm TeV}$&1&6&\\
$M_2/{\rm TeV}$&0.3&2&\\
$m_{{\tilde L}}=m_{{{\tilde e}^c}}/{\rm TeV}$&0.5&2&\\
$m_{{\tilde u}^c_3}/{\rm TeV}$&1&4&\\
$A_{t}/{\rm TeV}$&1&4&\\
\hline
\end{tabular}
\caption{Random scan parameters.}
\label{tab1}
\end{table}

To present numerical analysis, we random  scan the parameter space shown in Tab.~\ref{tab1}.  Considered that the light stop mass is easily ruled out by the experiment, we scan the parameter $m_{{\tilde u}^c_3}$ from 1 TeV.
Now the average measured mass of the Higgs boson is~\cite{PDG1}
\begin{eqnarray}
m_h=125.25\pm 0.17\: {\rm{GeV}},
\label{mh-exp}
\end{eqnarray}
where the accurate Higgs boson mass can  give stringent constraint on parameter space for the model. In our previous work~\cite{Zhang-MASS},  the Higgs boson masses in the $\mu\nu$SSM including the main two-loop radiative corrections are discussed.
Through the work, here the scanning results are constrained by the lightest Higgs boson mass with $124.68\,{\rm GeV}\leq m_{{h}} \leq125.52\:{\rm GeV}$, where a $3 \sigma$ experimental error is considered.
For the signal strengths of the light Higgs boson decay modes $h \rightarrow \gamma\gamma, \;WW^*, \;ZZ^*, \; b\bar b, \;\tau\bar\tau, \;\mu\bar\mu$, we adopt the averages of the results from PDG which reads~\cite{PDG1}
\begin{eqnarray}
&&\mu_{\gamma\gamma}^{exp}=1.11_{-0.09}^{+0.10},\quad
\mu_{WW^*}^{exp}=1.19\pm0.12,\quad
\mu_{ZZ^*}^{exp}=1.06\pm0.09,\nonumber\\
&&\mu_{b\bar b}^{exp}=1.04\pm0.13,\quad
\mu_{\tau\bar\tau}^{exp}=1.15_{-0.15}^{+0.16},\quad
\mu_{\mu\bar\mu}^{exp}=1.19\pm0.34.\label{hdecay}
\end{eqnarray}
Here  a $2 \sigma$ experimental error will be considered in  the scanning results, using our previous work~\cite{hrr} on the signal strengths of the Higgs boson decay channels $h\rightarrow\gamma\gamma$, $h\rightarrow VV^*$ ($V=Z,W$), and $h\rightarrow f\bar{f}$ ($f=b,\tau$) in the $\mu\nu$SSM.

There is a close similarity between the anomalous MDM of muon and the branching ratio of $\bar{B}\rightarrow X_s\gamma$ in the supersymmetric model~\cite{Martin:2001st}. They both get large  $\tan\beta$ enhancements from the down-fermion Yukawa couplings, ${Y_{{d_i}}} = {{{m_{{d_i}}}}}/{{{\upsilon_d}}} = {{{m_{{d_i}}\sqrt{\tan^2\beta+1}}}}/{{{\upsilon}}}$ and ${Y_{{e_i}}} = {{{m_{{l_i}}}}}/{{{\upsilon_d}}} = {{{m_{{l_i}}\sqrt{\tan^2\beta+1}}}}/{{{\upsilon}}}$ with $\upsilon=\sqrt{\upsilon_d^2+\upsilon_u^2}\simeq 174$ GeV. Combined with the experimental data from CLEO~\cite{ref-CLEO}, BELLE \cite{ref-BELLE1,ref-BELLE2} and BABAR~\cite{ref-BABAR1,ref-BABAR2,ref-BABAR3}, the current experimental value for the branching ratio of $\bar{B}\rightarrow X_s\gamma$ is~\cite{PDG1}
\begin{eqnarray}
{\rm{Br}}(\bar{B}\rightarrow X_s\gamma)=(3.49\pm0.19)\times10^{-4}.
\end{eqnarray}
Using our previous work about the rare decay $\bar{B}\rightarrow X_s\gamma$ in the $\mu\nu$SSM~\cite{ref-bsr}, the following results in the scanning are also constrained by  $2.92\times 10^{-4} \leq {\rm{Br}}(\bar{B}\rightarrow X_s\gamma) \leq 4.06\times 10^{-4}$, where a $3 \sigma$ experimental error is considered.

\subsection{The MDM of charged leptons}

\begin{figure}
\setlength{\unitlength}{1mm}
\centering
\begin{minipage}[c]{0.48\textwidth}
\includegraphics[width=2.8in]{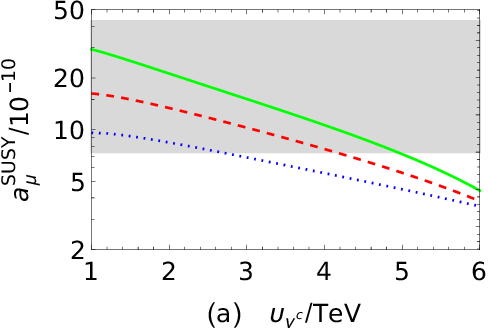}
\end{minipage}%
\begin{minipage}[c]{0.45\textwidth}
\includegraphics[width=2.8in]{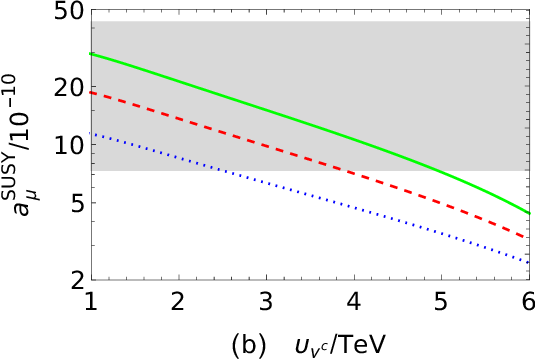}
\end{minipage}
\caption[]{ (Color online)  $a_\mu^{\rm{SUSY}}$ versus $\upsilon_{\nu^c}$ with different $M_2$ (a) and $\tan \beta$ (b), where the gray area denotes $\Delta a_\mu$ at $3.0\sigma$ given in Eq.~(\ref{MDM-exp}). }
\label{figvcau}
\end{figure}

Firstly, to illustrate clearly the cross-correlation of the model parameters, we plot $a_\mu^{\rm{SUSY}}$ varying with $\upsilon_{\nu^c}$ with different $M_2$ and $\tan \beta$ in Fig.~\ref{figvcau}, choosing $m_{{\tilde L}}=m_{{{\tilde e}^c}}=0.7$ TeV and $m_{{\tilde u}^c_3}=A_t=1$ TeV for simplicity. In Fig~\ref{figvcau}(a), the solid line denotes $M_2=0.3$ TeV, the dashed line denotes $M_2=1$ TeV, and the dotted line denotes $M_2=2$ TeV, with $\tan \beta =40$. The numerical results show that the muon anomalous MDM $a_\mu^{\rm{SUSY}}$  is decoupling with increasing  $\upsilon_{\nu^c}$ or  $M_2$, which can affect the masses of the charginos and neutralinos. In Fig~\ref{figvcau}(b), the solid line represents $\tan \beta =40$, the dashed line represents $\tan \beta =25$, and the dotted line represents $\tan \beta =15$, with $M_2=0.3$ TeV. Through Fig~\ref{figvcau}(b), we can see that the muon anomalous MDM $a_\mu^{\rm{SUSY}}$ get large  $\tan\beta$ enhancements, which is similar with that of the MSSM.

\begin{figure}
\setlength{\unitlength}{1mm}
\centering
\begin{minipage}[c]{0.48\textwidth}
\includegraphics[width=2.8in]{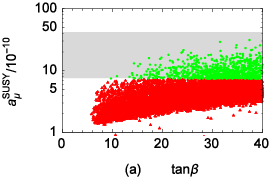}
\end{minipage}%
\begin{minipage}[c]{0.45\textwidth}
\includegraphics[width=2.8in]{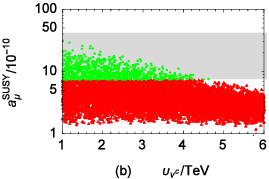}
\end{minipage}
\caption[]{ (Color online)  $a_\mu^{\rm{SUSY}}$ versus $\tan \beta$ (a) and $\upsilon_{\nu^c}$ (b), where the gray area denotes $\Delta a_\mu$ at $3.0\sigma$ given in Eq.~(\ref{MDM-exp}) and the red triangles are eliminated. }
\label{fig4}
\end{figure}

\begin{figure}
\setlength{\unitlength}{1mm}
\centering
\begin{minipage}[c]{0.48\textwidth}
\includegraphics[width=2.8in]{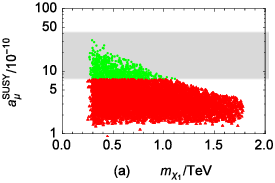}
\end{minipage}%
\begin{minipage}[c]{0.45\textwidth}
\includegraphics[width=2.8in]{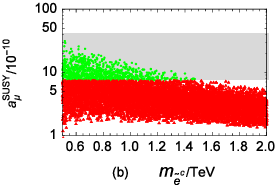}
\end{minipage}
\caption[]{ (Color online)  $a_\mu^{\rm{SUSY}}$ versus $m_{\chi_1}$ (a) and $m_{{{\tilde e}^c}}$ (b), where the gray area denotes $\Delta a_\mu$ at $3.0\sigma$ given in Eq.~(\ref{MDM-exp}) and the red triangles are eliminated. }
\label{fig5aum}
\end{figure}

Through random scanning the parameter space shown in Tab.~\ref{tab1}, we plot anomalous magnetic dipole moment of muon $a_\mu^{\rm{SUSY}}$ varying with the key parameters $\tan \beta$ (a) and $\upsilon_{\nu^c}$ (b) in Fig.~\ref{fig4}, where the gray area denotes $\Delta a_\mu$ at $3.0\sigma$ given in Eq.~(\ref{MDM-exp}).  Here, the red triangles are excluded by $\Delta a_\mu$ at $3.0\sigma$. The green points of the $\mu\nu$SSM are accord with $\Delta a_\mu$ at $3.0\sigma$, which can explain the current tension between the experimental measurement and the SM theoretical prediction for the muon anomalous MDM.

Fig.~\ref{fig4}(a) show that the muon anomalous MDM $a_\mu^{\rm{SUSY}}$ increases with increasing of the parameter  $\tan \beta$. One can find that a significant region of the parameter space is excluded by $\Delta a_\mu$ at $3.0\sigma$ in the small $\tan \beta$ region. Here, very small $\tan \beta$ region is also easily eliminated by the constraint of the 125 GeV Higgs boson mass.
The numerical results in Fig.~\ref{fig4}(b) depict that the muon anomalous MDM $a_\mu^{\rm{SUSY}}$ is decoupling with increasing of $\upsilon_{\nu^{c}}$. Due that $\upsilon_{\nu^{c}}$ can affect the masses of charginos and neutralinos. We can see that the value of the muon anomalous MDM $a_\mu^{\rm{SUSY}}$ in the $\mu\nu$SSM could explain the experimental muon anomalous MDM $\Delta a_{\mu}$ at $3.0\sigma$ shown in Eq.~(\ref{MDM-exp}),  when $\upsilon_{\nu^{c}}$ is small and $\tan \beta$ is large. Constrained by $\Delta a_{\mu}$ at $3.0\sigma$ shown in Eq.~(\ref{MDM-exp}), $\tan \beta<10$ or $\upsilon_{\nu^{c}}>5$ TeV will easily be  eliminated.

To see more clearly, we plot anomalous magnetic dipole moment of muon $a_\mu^{\rm{SUSY}}$ varying with
the lightest chargino mass $m_{\chi_1}$ in Fig.~\ref{fig5aum}(a), through scanning the parameter space shown in Tab.~\ref{tab1}. The results show that the contribution of the lightest chargino mass $m_{\chi_1}$ is roughly similar with the contribution of the parameter $\upsilon_{\nu^c}$. Because here  $\mu \equiv 3\lambda\upsilon_{\nu^{c}}$, where $\mu$ directly affect the masses of the charginos. In Fig.~\ref{fig5aum}(b), $a_\mu^{\rm{SUSY}}$ versus $m_{{{\tilde e}^c}}$ is also pictured. When $m_{{{\tilde e}^c}}$ is small, $a_\mu^{\rm{SUSY}}$ in the $\mu\nu$SSM could explain the $\Delta a_{\mu}$ at $3.0\sigma$. The variation trend of $a_\mu^{\rm{SUSY}}$ versus $m_{{{\tilde e}^c}}$ coincides with the decoupling theorem, due that $m_{{{\tilde e}^c}}$ directly affects the masses of the slepton. One can see that anomalous magnetic dipole moment of muon $a_\mu^{\rm{SUSY}}$ can reach $\Delta a_{\mu}$ at $3.0\sigma$ shown in Eq.~(\ref{MDM-exp}), when $m_{\chi_1}<1.1$ TeV and  $m_{{{\tilde e}^c}}<1.5$ TeV.

\begin{figure}
\setlength{\unitlength}{1mm}
\centering
\begin{minipage}[c]{0.48\textwidth}
\includegraphics[width=2.8in]{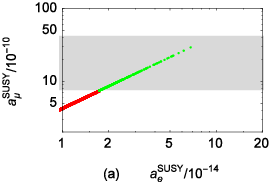}
\end{minipage}%
\begin{minipage}[c]{0.45\textwidth}
\includegraphics[width=2.8in]{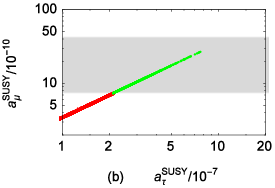}
\end{minipage}
\caption[]{ (Color online)  $a_\mu^{\rm{SUSY}}$ versus $a_e^{\rm{SUSY}}$ (a) and $a_\tau^{\rm{SUSY}}$ (b), where the gray area denotes $\Delta a_\mu$ at $3.0\sigma$ given in Eq.~(\ref{MDM-exp}) and the red triangles are eliminated by $\Delta a_\mu$ at $3.0\sigma$. }
\label{fig6}
\end{figure}

For the anomalous MDM of the electron and tau lepton, we also picture  $a_\mu^{\rm{SUSY}}$ versus $a_e^{\rm{SUSY}}$ (a) and $a_\tau^{\rm{SUSY}}$ (b) in Fig.~\ref{fig6}, where the green points are in agree with $\Delta a_\mu$ at $3.0\sigma$ given in Eq.~(\ref{MDM-exp}) and the red triangles are eliminated by that. Constrained by the updated discrepancy for  $\Delta a_\mu$ at $3.0\sigma$,  the anomalous MDM $a_e^{\rm{SUSY}}$ and $a_\tau^{\rm{SUSY}}$ in the $\mu\nu$SSM can be about $0.7\times10^{-13}$ and $0.8\times10^{-6}$, respectively.
The numerical results show that the ratio between the anomalous MDMs of the tau lepton and muon is about $2.8\times10^2$, which is in agree with $\frac{\bigtriangleup a_\tau}{\bigtriangleup a_\mu}\simeq m_\tau^2/m_\mu^2\simeq2.8\times10^2$. The ratio between the anomalous  MDMs of the muon and electron also is consistent with  $\frac{\bigtriangleup a_\mu}{\bigtriangleup a_e}\simeq m_\mu^2/m_e^2\simeq4.3\times10^4$.

\subsection{$h\rightarrow l_i \bar{l}_i$}

\begin{figure}
\setlength{\unitlength}{1mm}
\centering
\begin{minipage}[c]{0.5\textwidth}
\includegraphics[width=2.8in]{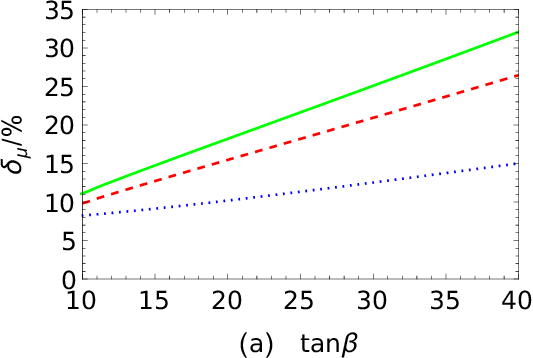}
\end{minipage}%
\begin{minipage}[c]{0.5\textwidth}
\includegraphics[width=2.85in]{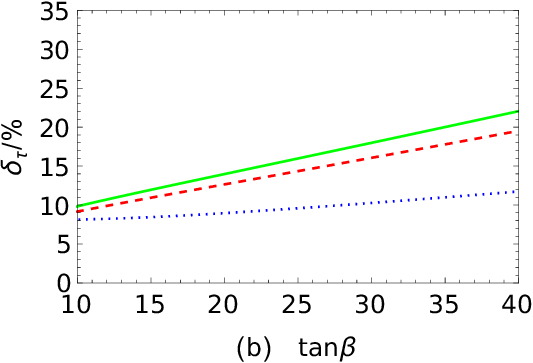}
\end{minipage}
\caption[]{(Color online)  The ratio $\delta_\mu$ (a) and  $\delta_\tau$ (b) versus the parameter   $\tan\beta$ with different $\upsilon_{\nu^c}$.}
\label{Figtbh}
\end{figure}

We define the physical quantity
\begin{eqnarray}
\delta_{l_i}\equiv {{\Gamma}_{\rm{NP}}(h\rightarrow {l_i}\bar{{l_i}})-{\Gamma}_{\rm{SM}}(h\rightarrow {l_i}\bar{{l_i}})\over {\Gamma}_{\rm{SM}}(h\rightarrow {l_i}\bar{{l_i}})},
\end{eqnarray}
to show the difference of the decay width of $h\rightarrow {l_i}\bar{{l_i}}$ of the $\mu\nu$SSM (${\Gamma}_{\rm{NP}}(h\rightarrow {l_i}\bar{{l_i}})$) and that of the SM (${\Gamma}_{\rm{SM}}(h\rightarrow {l_i}\bar{{l_i}})$), where ${l_i}=e,\mu,\tau$.
Firstly, to illustrate clearly the cross-correlation of the model parameters, we plot the ratio $\delta_\mu$ (a) and  $\delta_\tau$ (b) versus the parameter  $\tan\beta$ with different $\upsilon_{\nu^c}$ in Fig.~\ref{Figtbh}, taking $m_{{\tilde L}}=m_{{{\tilde e}^c}}=0.6$ TeV, $M_2=2$ TeV, $m_{{\tilde u}^c_3}=2$ TeV and $A_t=3$ TeV for simplicity. In Fig.~\ref{Figtbh}, the solid line denotes $\upsilon_{\nu^c}=6$ TeV, the dashed line denotes $\upsilon_{\nu^c}=3$ TeV, and the dotted line denotes $\upsilon_{\nu^c}=1$ TeV, respectively. The numerical results in Fig.~\ref{Figtbh} show that the ratio $\delta_\mu$ and  $\delta_\tau$  increase with increasing of $\tan\beta$ or $\upsilon_{\nu^{c}}$. The charged leptons Yukawa couplings get large $\tan\beta$ enhancements, with ${Y_{{e_i}}} = {{{m_{{l_i}}\sqrt{\tan^2\beta+1}}}}/{{{\upsilon}}}$.

\begin{figure}
\setlength{\unitlength}{1mm}
\centering
\begin{minipage}[c]{0.5\textwidth}
\includegraphics[width=2.8in]{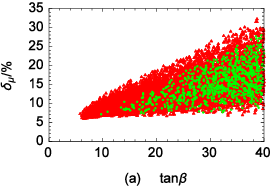}
\end{minipage}%
\begin{minipage}[c]{0.5\textwidth}
\includegraphics[width=2.85in]{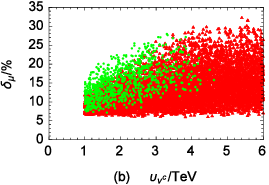}
\end{minipage}
\caption[]{(Color online)  The ratio $\delta_\mu$ versus the parameter   $\tan\beta$ (a) and $\upsilon_{\nu^{c}}$ (b).}
\label{Fig-D2}
\end{figure}

\begin{figure}
\setlength{\unitlength}{1mm}
\centering
\begin{minipage}[c]{0.5\textwidth}
\includegraphics[width=2.8in]{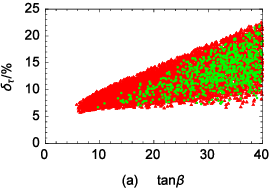}
\end{minipage}%
\begin{minipage}[c]{0.5\textwidth}
\includegraphics[width=2.85in]{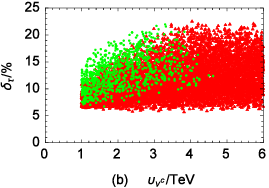}
\end{minipage}
\caption[]{(Color online)  The ratio $\delta_\tau$ versus the parameter  $\tan\beta$ (a) and $\upsilon_{\nu^{c}}$ (b).}
\label{Fig-D3-tbvc}
\end{figure}

Through scanning in Tab.~\ref{tab1}, we plot Figs.~\ref{Fig-D2}-\ref{Fig-D3-tbvc}, where the green dots are the corresponding physical quantity's values of the remaining parameters after being constrained by the muon anomalous MDM $a_{\mu}^{\rm{SUSY}}$ in the $\mu\nu$SSM with $7.4\times 10^{-10} \leq a_{\mu}^{\rm{SUSY}} \leq 42.8\times 10^{-10}$ considered a $3 \sigma$ experimental error. The red triangles are ruled out by the muon anomalous MDM with $a_{\mu}^{\rm{SUSY}}> 42.8\times 10^{-10}$ and $a_{\mu}^{\rm{SUSY}} < 7.4\times 10^{-10}$.

In Fig.~\ref{Fig-D2}, we plot the ratio $\delta_\mu$ varying with the parameter $\tan\beta$ (a) and $\upsilon_{\nu^{c}}$ (b). We can see that the ratio $\delta_\mu$ will increase with increasing of $\tan\beta$ in Fig.~\ref{Fig-D2}(a). The ratio $\delta_\mu$ can be close to 30$\%$ when the parameter $\tan\beta$ is large, Constrained by  $\Delta a_\mu$ at $3.0\sigma$.
In Fig.~\ref{Fig-D2}(b) shows that the ratio $\delta_\mu$ is non-decoupling with increasing  $\upsilon_{\nu^{c}}$. The maximum of the ratio $\delta_\mu$ can be around 15$\%$ as $\upsilon_{\nu^c}$ is about 1 TeV and close to 30$\%$ as $\upsilon_{\nu^c}$ is about 3 TeV.
In the $\mu\nu$SSM, the parameter $\upsilon_{\nu^{c}}$ leads to the mixing of the neutral components of the Higgs doublets with the sneutrinos. The mixing affects the lightest Higgs boson mass and the Higgs couplings, which is different from the SM.

In addition, we plot the ratio $\delta_\tau$ varying with the parameter $\tan\beta$ and $\upsilon_{\nu^{c}}$ in Fig.~\ref{Fig-D3-tbvc}, which the variation trend is similar with that of the ratio $\delta_\mu$. The numerical results show that, constrained by the experimental value of the muon anomalous MDM, the ratio $\delta_\tau$ can be about $20\%$, when $\tan \beta$ is about 40 and $\upsilon_{\nu^c}$ is around 3 TeV.

\section{Summary\label{sec-Sum}}

Considered that the new experimental average for the muon anomalous MDM increases the tension between experiment and SM prediction to 4.2$\sigma$, we analyse the muon anomalous MDM at two-loop level in the $\mu\nu$SSM. The numerical results show that the $\mu\nu$SSM can explain the current tension between the experimental measurement and the SM theoretical prediction for the muon anomalous MDM, constrained by the 125 GeV Higgs boson mass and decays, the rare decay $\bar{B}\rightarrow X_s\gamma$ and so on. The new experimental average of the muon anomalous MDM considered a $3 \sigma$ also give a strictly constraint for the parameter space of the $\mu\nu$SSM, which constrain that $\tan \beta>10$, $m_{{{\tilde e}^c}}<1.5$ TeV and $\upsilon_{\nu^{c}}<5$ TeV with $\lambda=0.1$. And the anomalous MDM of tau lepton and electron in the $\mu\nu$SSM can reach about $0.7\times10^{-13}$ and $0.8\times10^{-6}$ respectively, constrained by the new experimental average of the muon anomalous MDM at $3.0\sigma$.

An upgrade to the Muon g-2 experiment at Fermilab and another experiment at J-PARC \cite{J-PARC} will measure the muon anomalous magnetic dipole moment with higher precision, which may reach a 5$\sigma$ deviation from the SM, constituting an augury for new physics beyond the SM. And the anomalous MDM of tau lepton and electron whether deviate from the SM prediction will be given more accurately, with the development of experiment in the future.

Considered that the ATLAS and CMS Collaborations measured the 125 GeV Higgs boson decay to a pair of muons $h\rightarrow \mu \bar{\mu}$ recently, we also investigate the 125 GeV Higgs boson decay $h\rightarrow \mu \bar{\mu}$ at one-loop level in the $\mu\nu$SSM. Compared to the SM prediction, the decay width of $h\rightarrow \mu \bar{\mu}$ and $h\rightarrow \tau \bar{\tau}$ in the $\mu\nu$SSM can boost up about 30\% and 20\%, considering the constraint from the muon anomalous magnetic dipole moment. In the $\mu\nu$SSM, the mixing of the neutral components of the Higgs doublets with the sneutrinos affects the lightest Higgs boson mass and the Higgs couplings, which can contribute to the Higgs boson decay. In the future, high luminosity or high energy large colliders~\cite{ref-100pp,ref-HL,ref-CEPC,ref-ILC} will detect the Higgs boson decay $h\rightarrow \mu \bar{\mu}$ and $h\rightarrow \tau \bar{\tau}$  with high precision, which may see the indication of new physics.

\begin{acknowledgments}
\indent\indent
The work has been supported by the National Natural Science Foundation of China (NNSFC) with Grants No. 11705045, No. 11535002, No. 12075074, the youth top-notch talent support program of the Hebei Province,  and Midwest Universities Comprehensive Strength Promotion project.
\end{acknowledgments}

\end{document}